# A three-dimensional MR-STAT protocol for high-resolution multi-parametric quantitative MRI

Hongyan Liu[1], Oscar van der Heide[1], Edwin Versteeg[1], Martijn Froeling[2], Miha Fuderer[1], Fei Xu[1], Cornelis A.T. van den Berg[1], and Alessandro Sbrizzi[1]

[1] Computational Imaging Group for MRI Therapy & Diagnostics, Department of Radiotherapy, University Medical Center Utrecht, Utrecht, Netherlands

[2] Department of Radiology, Imaging Division, University Medical Center Utrecht, Utrecht, Netherlands

**Corresponding author:** Hongyan Liu, h.liu@umcutrecht.nl

**Abstract summary.**

Magnetic Resonance Spin Tomography in Time-Domain (MR-STAT) is a multiparametric quantitative MR framework, which allows for simultaneously acquiring quantitative tissue parameters such as T1, T2 and proton density from one single short scan. A typical 2D MR-STAT acquisition uses a gradient-spoiled, gradient-echo sequence with a slowly varying RF flip-angle train and Cartesian readouts, and the quantitative tissue maps are reconstructed by an iterative, model-based optimization algorithm. In this work, we design a 3D MR-STAT framework based on previous 2D work, in order to achieve better image SNR, higher though-plan resolution and better tissue characterization.

Specifically, we design a 7-minute, high-resolution 3D MR-STAT sequence, and the corresponding two-step reconstruction algorithm for the large-scale dataset. To reduce the long acquisition time, Cartesian undersampling strategies such as SENSE are adopted in our transient-state quantitative framework. To reduce the computational burden, a data splitting scheme is designed for decoupling the 3D reconstruction problem into independent 2D reconstructions. The proposed 3D framework is validated by numerical simulations, phantom experiments and in-vivo experiments. High-quality knee quantitative maps with 0.8 x 0.8 x 1.5mm3 resolution and bilateral lower leg maps with 1.6mm isotropic resolution can be acquired using the proposed 7-minute acquisition sequence and the 3-minute-per-slice decoupled reconstruction algorithm. The proposed 3D MR-STAT framework could have wide clinical applications in the future.

1. Introduction

Quantitative MRI (qMRI) techniques measure parameter maps such as relaxation times ($T_1$, $T_2$, $T_2^*$) and proton density ($\rho$). Compared to conventional qualitative images, quantitative tissue maps reveal tissue properties which are independent of system conditions and human interpretation, and have shown promising potential for better clinical diagnosis of various pathological changes [1]–[5]. During the past decade, several multiparametric qMRI techniques have been developed to obtain multiple tissue parameters from one single scan, to accelerate the acquisitions, and eliminate motion between separate acquisitions. Examples of such techniques include MR Fingerprinting [6], [7], 3D-QALAS (a 3D interleaved Look-Locker sequence with T2 preparation pulse) [8], [9], QTI (quantitative transient-state imaging) [10], [11], MR Multitasking [12], [13] and MR-STAT [14], [15].

Magnetic Resonance Spin Tomography in Time-Domain (MR-STAT) [14] is a multiparametric qMRI framework, which allows for simultaneously obtaining $T_1$, $T_2$ and $\rho$ from one single short scan. A typical two-dimensional MR-STAT acquisition uses a gradient-spoiled, gradient-echo sequence with a slowly varying flip-angle train and Cartesian readouts [16]. Subsequently, the spatial parameter maps are reconstructed by solving a model-based large-scale nonlinear optimization problem directly from time-domain data. A five-minute whole-brain protocol that contains 30 slices of 2D Cartesian MR-STAT acquisitions has been developed and assessed in a clinical study, providing diagnostic accuracy close to conventional multi-scan contrast-weighted images [17].

The current MR-STAT 2D acquisitions [17] use a resolution of 1 x 1 x 3 mm$^3$, and still suffer from a relatively low signal-to-noise ratio (SNR). Three-dimensional MR imaging can provide image volumes with high resolutions in all three directions as well as higher SNRs compared to 2D protocols. Therefore, the main purpose of this work is to extend the current 2D Cartesian MR-STAT sequence to 3D acquisitions, to achieve high-quality, 3D quantitative image volumes for better visualization and tissue quantification.

In general, the two main challenges for developing a 3D MR-STAT framework are (i) the long acquisition time for high-resolution 3D coverage and (ii) the computational burden for accurate model-based reconstructions of large 3D datasets. To reduce the long acquisition time, we adopt Cartesian undersampling strategies such as SENSE and CAIPIRINHA [18] and build them in our transient-state quantitative framework. To reduce the computational burden, the sequence is also designed to make a data splitting scheme possible, such that the 3D reconstruction problem can be decoupled into multiple 2D reconstructions. The decoupled reconstruction strategy makes the 3D reconstruction possible on a Desktop PC with a runtime comparable to 2D reconstructions, approximately 3 minutes per slice.

The new 3D MR-STAT framework was validated by (i) numerical experiments, (ii) gel phantom measurements and (iii) retrospective and prospective undersampled in-vivo scans.

2. Theory

2.1    Three-dimensional sequence design

The current 2D MR-STAT sequence is a Cartesian gradient-spoiled, gradient echo sequence with varying flip-angles preceded by a non-selective inversion pulse. Let $kx$ be the readout direction and $ky$ be the phase-encoding direction in the k-space. For 2D acquisitions, a linear, Cartesian sampling strategy is employed, and a total number of $N_k$ k-space data are fully sampled. Typically, 4 <$N_k$ < 10. Let $N_{ky}$ be the number of fully-sampled phase-encoding lines per k-space, which equals to $N_y$, the image size in the $y$ direction. The total number of excitations is thus $N_{ky} * N_k$.

Let $\boldsymbol{d}(t)$ be the signal measured from an MR-STAT acquisition, the MR-STAT reconstruction solves the following nonlinear optimization problem [14]

$$\boldsymbol{\alpha}^* = \arg\min_{\boldsymbol{\alpha}} \int \|\boldsymbol{d}(t) - \boldsymbol{s}(\boldsymbol{\alpha}, t)\|_2^2 dt, \quad (1)$$

where $\boldsymbol{\alpha}$ encapsulates all spatial parameter maps and $s(\boldsymbol{\alpha}, t)$ is the volumetric integral of the modeled local transverse magnetization given $\boldsymbol{\alpha}$.

Reconstructing a whole 3D MR-STAT data by directly solving (1) with gradient-based iterative methods can be extremely challenging due to the excessive demands on reconstruction time and computer memory [16], [19]. Therefore, we designed our 3D framework using segmented readouts based on the current 2D acquisition, which can be beneficial for later simplification of the data reconstruction step.

A 3D MR-STAT acquisition can be realized by concatenating multiple 2D acquisitions by:

(a) introducing a second phase-encoding direction, $kz$, and
(b) repeating the 2D MR-STAT sequence for each $kz$ index as a segment of the 3D sequence, and
(c) replacing the 2D slice-selective RF excitation pulse with a 3D slab-selective excitation pulse.

Let $N_{kz}$ be the number of 3D segment repetitions for fully-sampled acquisitions, and $N_{kz}$ equals to $N_z$, where $N_z$ is the image size in the $z$ direction. For a 3D fully-sampled acquisition, the total scan time is $N_{kz} \cdot N_{ky} \cdot N_k \cdot TR$. Given as an example, $N_{ky} = 224, N_{kz} = 128, N_k = 5, TR = 10ms$, the estimated scan time would be about 24 minutes. This prolonged scan time motivates the necessity of accelerating the acquisition with parallel imaging techniques.

A schematic illustration of a 3D cartesian MR-STAT sequence with two different undersampling strategies is shown in **Figure 1(a)**. The first undersampling strategy is represented by the cross-hatched and solid points, which show an undersampling factor of 2 along the $ky$ phase encoding direction only (the $kz$ direction is fully sampled); this is the standard SENSE undersampling pattern, and halves the length of 3D segments. Additional acceleration can be obtained by sampling only the solid points following the yellow trajectory; a total undersampling factor of 4 is thus achieved in the $ky - kz$ phase encoding plane, which resembles a 2D CAIPIRINHA pattern [18].

3D RF pulse excites the whole 3D volume for every repetition and a waiting time $T_w$ equal to a multiple of the longest $T_1$ should be inserted to allow all spins to fully recover to the equilibrium state. Such a long pause would greatly slow down the acquisition. However, an initial equilibrium state is not mandatory at the beginning of the 3D segments. Employing shorter values for $T_w$, the temporal MR signal response achieves a steady state ("hyper-steady state") after the first few repetitions of the 3D segment [20], [21]. Examples of the simulated MR signals after multiple repetitions of 3D segments with different $T_w$ values are shown in **Figure 2**. This shows that the MR signals with the same flip-angle index from all segment repetitions except for the first three are in the same temporal transient-state. The "hyper steady-state" condition allows us to simulate only the first few segments instead of the whole 3D sequence, leading to substantial computational advantages; in addition, the coherence of excitations in the $kz$ direction (repetitions) makes the data decoupling strategy possible, as we will show in the following section.

## 2.2  Reconstruction strategy

The 3D sequence presented in the previous section allows us to design a fast and memory-efficient 3D MR-STAT reconstruction workflow, in which we develop a decoupling strategy to split the whole reconstruction into two substeps (see **Figure 3**):

> **Substep 1**: a CG (conjugate gradient)-SENSE reconstruction is run for each of the $N_k$ undersampled 3D k-space datasets to acquire the "fully-sampled" k-space data. Once the fully-sampled dataset is recovered, this can be decoupled by one-dimensional FFT in the $kz$ phase-encoding direction into multiple slices of 2D MR-STAT k-space data;

> **Substep 2**: quantitative maps are reconstructed using the decoupled 2D k-space data by running slice-by-slice 2D MR-STAT reconstructions. The 2D reconstructions can be performed as proposed in [22].

**Figure 3** shows the flowchart of the 3D MR-STAT reconstruction using the decoupling strategy. In Substep 1, the multi-channel undersampled 3D MR-STAT data are firstly transformed into the image space allowing for a SENSE reconstruction to fill in the missing data samples. The SENSE reconstructed images are then transformed to 2D $kx - ky$ space to get the fully-sampled 2D MR-STAT k-space data. Note that the SENSE reconstruction here is run on MR-STAT FFT images transformed from transient-state k-space data, and the k-space data samples with different $ky$ indexes are in different transient states. This suggests that the SENSE reconstruction might not be perfect. However, when the flip-angle trains and the corresponding signal responses are sufficiently smooth, the SENSE reconstruction can still recover the missing samples of the transient-state data with a good approximation which we will demonstrate using numerical simulations. In Substep 2, an accelerated MR-STAT algorithm is used for the 2D slice-by-slice reconstruction [22], which realizes fast and memory-efficient tissue parameter mapping by applying a neural network model for signal computation and incorporating an alternating direction method of multipliers (ADMM) [22].

In Substep 2, an accelerated MR-STAT algorithm is used for the 2D slice-by-slice reconstruction [22], which realizes fast and memory-efficient tissue parameter mapping by applying a neural network model for signal computation and incorporating an alternating direction method of multipliers (ADMM).

## 3.  Methods

To test the proposed sequence design and reconstruction strategy, we used (i) numerical simulations, on a 3D digital brain phantom (ii) measurements on a gel phantom and (iii) retrospective and prospective undersampled in-vivo scans of two healthy volunteers after acquiring written informed consent.

The two-step reconstruction strategy explained in Section 2.2 (see also **Figure 3**) was used for reconstructing simulated and measured 3D MR-STAT data. All reconstructions were run on an 8-core desktop PC with 3.7GHz CPUs. In Substep 1, 20 iterations of CG-SENSE were run for each of the five 3D k-space data [23]. In Substep 2 for the 2D MR-STAT reconstructions, 15 iterations of the ADMM presented in were run for each slice.

### 3.1 Numerical simulations

A simulation study was first performed to validate the 3D MR-STAT framework. A digital brain phantom [24] with a matrix size of $N_x \times N_y \times N_z = 184 \times 184 \times 128$ and a resolution of $1.2mm^3$ was used to generate the synthetic data. The numbers of fully-sampled phase-encoding lines are $N_{ky} = N_y = 184$, $N_{kz} = N_z = 128$. A total undersampling factor of 4, (2-fold in $ky$ direction and 2-fold in $kz$ direction), was applied in the numerical experiment, as shown in **Figure 1(a)**. A RF flip-angle train with four sine-square lobes over five k-spaces was used for a 3D segment (see Sequence A in **Figure 1(b)**); the flip angle values are given by

$$\text{FA}(n) = 70 * \sin^2\left(\frac{4\pi*n}{N_{ky}/2*N_k}\right) [\text{degree}], n = 1,2,\dots,N_{ky}/2*N_k. \quad (2)$$

The length of each 3D segment was $N_{ky}/2 * N_k = 460$ excitations, and the number of repetitions for the undersampled sequence was $N_{kz}/2 = 64$. An RF excitation pulse with a perfect slice profile was used for the simulation. To get good $T_1$ and $T_2$ encoding, an inter-repetition pause $T_w = 1.75s$ and $TR = 10.4ms$ were applied. The total scan time was 6 minutes 57 seconds.

To simulate the synthetic data, tissue parameters were assigned to every tissue type of the digital phantom according to the literature. Realistic receive coil maps were obtained by registering a 13-channel receive array from an in-vivo volunteer scan to the digital brain phantom. The resulting maps were used for simulating multi-channel MR-STAT data[25], [26]. A homogeneous RF excitation field was assumed for the whole volume. Complex Gaussian noise was added to each channel of the simulated k-space data with an SNR of 40.

The SENSE reconstruction (Substep 1) of the transient-state data in our proposed splitting strategy relies on the assumption that the signal variation is very smooth, which is true if the flip angle varies slowly in time. We expect that the Substep 1 will be less accurate for faster variations of the signal over the k-space coverage. To investigate how a faster oscillating flip-angle train can affect the accuracy of the transient-state SENSE reconstruction and the corresponding quantitative maps after MR-STAT reconstructions (Substep 2), we simulated two additional numerical experiments with faster oscillating RF flip-angle trains: with 8 or 24 sine-square lobes instead of the standard 4 (see Sequence B and C in **Figure 1(b)**).

**3.2 MR experiments**

The 3D MR stat sequence was implemented on a 3T MRI system (Philips). The sequence implementation used the standard 4 sine-square lobes flip-angle train and allowed for either ky undersampling pattern (undersampling factor of 2 in ky direction, i.e. SENSE pattern) or ky-kz undersampling pattern (undersampling factor of 2 in both ky and kz directions, i.e. CAIPIRINHA pattern). For the sequence, an asymmetric sinc-gauss RF excitation pulse was used to excite the whole 3D volume, and an oversampling factor of 1.28 along the slab selective direction was used to avoid aliasing. The imperfect slab profile was considered in the EPG-based signal model computation during the reconstruction to acquire accurate quantitative maps.

Coil sensitivity maps required for SENSE reconstruction were acquired by an additional low-resolution reference scan and reconstructed by the ESPIRiT method [27]. Pre-whitening was executed for multi-channel raw data before the SENSE reconstruction to reduce the inter-channel noise correlation. A 30-second low-resolution, multi-slice B1 dream sequence was run for acquiring $B_1^+$ maps[28]. The acquired $B_1^+$ maps had a voxel size approximately 27 times larger than the voxel size from MR-STAT acquisitions and were interpolated to the same resolution to be used for the MR-STAT reconstructions.

Using this sequence we performed 3 MR experiments on (i) a gel phantom (ii) a knee (iii) lower legs which are summarized in **Table 1**. General sequence parameters were: $TR = 10.4 ms$, $TE = 4.2 ms$, waiting time between repetitions $T_w = 1.75 s$, readout bandwidth $BW = 44.65 kHz$.

### 3.2.1 Gel phantom experiment

For the gel phantom experiments, 11 gel-vials with different relaxation properties (TO5, Eurospin II test system, Scotland) were scanned using the 3D MR-STAT sequence with the $ky$ undersampling (R=2) pattern, which took 14 min. The data were retrospectively undersampled in the $kz$ direction by another factor of 2 to realize $ky - kz$ undersampling pattern (R=4). Additionaly, for SNR comparison, a central slice of the phantom was scanned with the 2D MR-STAT sequence. The 2D MR-STAT sequence had the same in-plane resolution of $1.2 \times 1.2 mm^2$, and a thicker slice of 3 mm. The same RF flip-angle pattern with 4 sine-square lobes was used and 5 full k-space data was sampled with a total number of RF excitations of 920. For validation Gold standard values were also acquired for the central slice of the gel tube phantom, using an inversion-recovery single spin-echo sequence with inversion times of [50, 100, 150, 350, 550, 850, 1250] ms for $T_1$ mapping, and a single-echo spin-echo sequence with echo times of [8, 28, 48, 88, 138, 188] ms for $T_2$ mapping.

For validation purpose, gold standard values were also acquired for the central slice of the gel tube phantom, using an inversion-recovery single spin-echo sequence with inversion times of [50, 100, 150, 350, 550, 850, 1250] ms for $T_1$ mapping, and a single-echo spin-echo sequence with echo times of [8, 28, 48, 88, 138, 188] ms for $T_2$ mapping.

### 3.2.2 in-vivo experiments

For the in vivo scans of the knee and lower leg, we used the same 3D MR STAT sequence as for the gel phantom. The knee acquisition used the $ky$ undersampling (R=2) pattern with additional retrospective undersampling by a factor 2, and the lower leg experiment used the prospective $ky - kz$ undersampling pattern (R=4), The knee $ky$ undersampling sequence took 14 minutes, and the leg $ky - kz$ undersampling sequence took 7 minutes.

To evaluate the $T_1$ and $T_2$ values in the in vivo data we used ROI analysis which were drawn on the proton density maps obtained from the reconstruction. For the knee results, representative ROIs including knee cartilage, bone marrow, muscle and subcutaneous fat were plot to compute quantitative values. For both legs, muscle segmentation is manually conducted by an experienced researcher on the reconstructed proton density maps, and the mean and std values of $T_1$ and $T_2$ for different muscles (anterior tibialis, extensor digitorum longus, soleus, gastrocnemius (lateral and medial head) and tibia bone marrow.

## 4. Results

### 4.1 Numerical simulations

The reconstruction results of the central slice for each substep of the reconstruction method and the Mean absolute percentage error (MAPE) values are shown in **Figure 4**. In Figure 4(a), the 2D MR-STAT data are shown for ease of comparison in the image domain by computing the Fourier transform of the k-space data for each of the 5 3D segments. These 2D FFT images encapsulate the transient-state signals, so they are not conventional steady-state $T_1$ / $T_2$-weighted images. Additionally, they are not the same for experiments using different flip-angle trains as is shown in rows.

Figure 4 demonstrates that when an RF flip-angle train oscillates faster, the two-step reconstruction strategy for undersampled 3D MR-STAT reconstruction yields larger reconstruction errors and aliasing artefacts in both substeps. **Figure S1** (Supplementary material) also shows similar reconstruction results as in Figure 4(b), but with fully-sampled 3D MR-STAT data, showing that quantitative maps can still be reconstructed with faster oscillating flip-angle trains. Results from **Figure 4(b)** and **Figure S1** show that using faster oscillating flip-angle trains leads to larger errors in the SENSE-based reconstruction (Substep 1), and therefore limits the possibility of running under-sampled acquisitions.

### 4.2 Gel phantom results

**Figure 5(a)** shows the reconstructed $T_1$ and $T_2$ maps for gel phantoms using retrospectively undersampled (R=4) 3D MR-STAT data, and the $B_1^+$ map from the B1 dream sequence used in MR-STAT reconstructions for B1 correction. **Figure 5(b)** shows the mean $T_1$ and $T_2$ values per tube computed from fully-sampled 2D MR-STAT results and 3D MR-STAT results (prospectively undersampling with R = 2 and retrospectively undersampling with R= 4). With $B_1^+$ correction, all MR-STAT results are in good agreement with the gold standard measurements.

To compare the SNR for 2D and 3D results, we define the sequence SNR efficiency by the $T_{1,2}$ -signal-to-noise ratio divided by the square root of the scan time per slice and the slice thickness [6]. **Figure 5(c)** plots the sequence efficiencies for 2D and 3D MR-STAT results. It can be observed that the 4-fold accelerated 3D MR-STAT has more than 400% higher $T_1$ and $T_2$ sequence efficiency compared to 2D fully-sampled MR-STAT.

### 4.3 In-vivo experimental results

**Figure 6** shows representative **(a)** sagittal and **(b)** transverse slices of the quantitative maps for the in-vivo knee scan. Results using both the 14-minute scan and the retrospectively undersampled 7-minute scan are shown. Mean and std values of $T_1$ and $T_2$ in ROI regions are calculated and reported in **Figure 6(c)**. Little difference is observed between the 7-minute results and 14-minute results in terms of the mean and std values in different ROIs.

**Figure 7** shows representative **(a)** coronal and **(b)** transverse slices of the quantitative maps for the bilateral lower leg scan. Results using the prospectively 4-fold undersampled 7-minute scan are shown. ROI analysis of the lower leg data are reported in (Figure 7(c)). Similar $T_1$ and $T_2$ values are observed for different muscles regions in left and right lower leg, except for the anterior tibialis; we observed that relatively low $B_1^+$ values (40%-60% of nominal $B_1^+$) are measured in the left anterior tibialis region, leading to lower mean $T_1$ (-13.3%) and $T_2$ (-14.0%) values compared to the right anterior tibialis region.

### 5. Discussion

In this paper, we present a 3D MR-STAT sequence with a two-step reconstruction algorithm, which can be used for acquiring volumetric quantitative maps with high image quality. A 7-minute 3D sequence with an undersampling factor of 4 was designed, implemented and demonstrated by reconstructing $T_1$, $T_2$ and proton density of a knee (0.8 x 0.8 x 1.5 mm³ resolution) and lower legs (1.6 x 1.6 x 1.6 mm³ resolution). In the reconstruction algorithm, Substep 1 recovers fully-sampled 2D MR-STAT data from 3D undersampled data by SENSE reconstruction which can be done for the transient-state MR-STAT data because (1) a smooth flip-angle train is used such that the transient-state signal is slowly varying and ky undersampling can be applied and (2) the hyper steady-state condition is reached such that repetitions of the 3D segments have the same signal response and thus kz undersampling can be applied. The SENSE reconstruction at Substep 1 is relatively fast and takes approximately 10 minutes. In Substep 2, the slice-by-slice 2D MR-STAT reconstruction takes approximately 5 hours in total (3 minutes per slice) on a Desktop PC. Since the 2D slices can be reconstructed separately, parallelization can easily be applied to accelerate the total reconstruction time.

As demonstrated by the phantom experiment, the current 3D MR-STAT sequence can be used for the reconstruction of accurate $T_1$ and $T_2$ maps when a separately acquired $B_1^+$ map is available. For the lower leg experiment, quantitative values are consistent with previously published results [29], [30]. However, a relatively large range of $B_1^+$ values (ranging from 40% to 150% of the nominal $B_1^+$) are measured within the 3D field-of-view, and lower $T_1$ and $T_2$ values are observed in the low $B_1^+$ (around 50% of nominal $B_1^+$) regions. This $T_1$ and $T_2$ inaccuracy may be caused by: (1) $B_1^+$ mapping inaccuracy [28], [31], especially in low $B_1^+$ regions, (2) the limited $T_1$ and $T_2$ encoding capacity of the MR-STAT sequence for extremely low $B_1^+$ values. In order to improve the accuracy for large range of $B_1^+$ values, MR-STAT sequences which are less sensitive to RF field variations should be developed in the future [32], [33]. Furthermore, the difference between std values from 14-minute data and 7-minute knee data is smaller than the difference observed in the phantom results. This suggests that the std measured in-vivo is dominated by real anatomical structures within the segmented areas or by model imperfections (such as chemical shift effects), rather than simple thermal noise. The measured mean $T_1$ and $T_2$ values of different tissue types mostly fall into the reference ranges from [25], [34], [35], except for the $T_2$ values for bone marrow and fat, which are longer than the literature values.

The current 3D MR-STAT framework only models the single-compartment signal and thus does not take into account chemical shift effects. The reconstructed water $T_2$ values (178-212ms) for fat and bone marrow regions in both the knee and lower legs experiments are higher than values reported in the literature (40-160ms) [25], which were measured by multi-echo spin-echo sequences. More accurate fat $T_2$ values can be acquired by modeling the spectral components of the fat signal in the physics model, and by separating water and fat components with multi-echo DIXON-type acquisitions applied to the MR-STAT sequences [36]. This will be investigated in future work. Although there have been existing 3D quantitative protocols faster than our presented 7-minute sequence, for example, 3D EPTI [37] (a 3-minute scan at 1-mm isotropic resolution with whole-brain coverage), optimized multi-axis spiral-projection MRF [7] (1-mm isotropic whole-brain quantitative mapping in 2 minutes), we believe that there is additional room to further shorten the 3D MR-STAT acquisition. One way to accelerate the current sequence is to optimize the sequence parameters, such as $TR$ and waiting time $T_w$ based on Cramer-Rao-based methods. The fast BLAKJac analysis technique [38] allows for sequence parameter optimization for a given phase-encoding pattern, and could achieve optimal tissue parameter encodings. Using the BLAKJac technique to optimize the 3D MR-STAT sequence could potentially reduce the scan time without sacrificing image quality.

To further reduce the scan time for higher-resolution experiments, another possible solution is to apply more efficient undersampling strategies, such as compressed sensing[39], [40], to replace the current Cartesian sampling pattern in the phase-encoding directions. To reconstruct highly undersampled data, more advanced reconstruction methods, for example, the low-rank tensor-based signal modelling [41]–[43], and the model-based deep-learning reconstructions [44]–[48], will also be explored for future 3D MR-STAT algorithm design.

Our 3D MR-STAT sequence is an unbalanced gradient-echo sequence with Cartesian readouts. This type of sequence is known to be highly sensitive to motion [49] as well as flow in cerebrospinal fluid and blood vessels. We've also tested our 3D MR-STAT sequence for brain scans and we observed that artefacts caused by the CSF flow are more dominant in the 3D results compared to 2D MR-STAT; these artefacts are mostly visible along the $kz$ (feet-head) phase-encoding direction, as shown in Supplementary material (**Figure S2**). Flow rates in CSF regions change during the cardiac cycle due to the pulsatile nature of the flow, leading to ghosting artefacts along the phase-encoding gradient direction[50], [51]. Similar problems have also been observed by other quantitative MRI techniques, especially when using Cartesian MR Fingerprinting sequences [52]–[54]. To modify our 3D MR-STAT sequence for brain scans, one possible solution is to apply incoherent (pseudo-random) sampling strategies; ideally, the CSF flow-induced phase deviation would be reconstructed as random noise instead of ghosting artefacts [55]. Standard denoising techniques could thus be applied to further suppress these artefacts during the MR-STAT reconstruction.

## 6. Conclusion

A three-dimensional MR-STAT technique was developed using a repetitive flip-angle train adopting a CAIPIRINHA undersampling scheme. The reconstruction is performed by splitting the 3D problem into multiple slice-by-slice 2D MR-STAT reconstructions. The proposed method can obtain high-quality knee quantitative maps (T1, T2 and $\rho$) with 0.8 x 0.8 x 1.5mm$^3$ resolution and bilateral lower leg maps with 1.6mm isotropic resolution using a 7-minute sequence.

**FIGURES**

**Figure 1.** (a) Schematic Example of a 3D cartesian MR-STAT sequence. The dashed and solid points show an undersampling pattern in one phase-encoding (ky) direction, and only the solid points following the yellow trajectory show a CAIPIRINHA undersampling pattern in the 2D phase-encoding (ky-kz) plane. (b) Example RF flip-angle trains with 4, 8 or 24 sine-square lobes over 5 k-spaces. The 4-lobe flip-angle train (Sequence A) is used for phantom and in-vivo experiments.

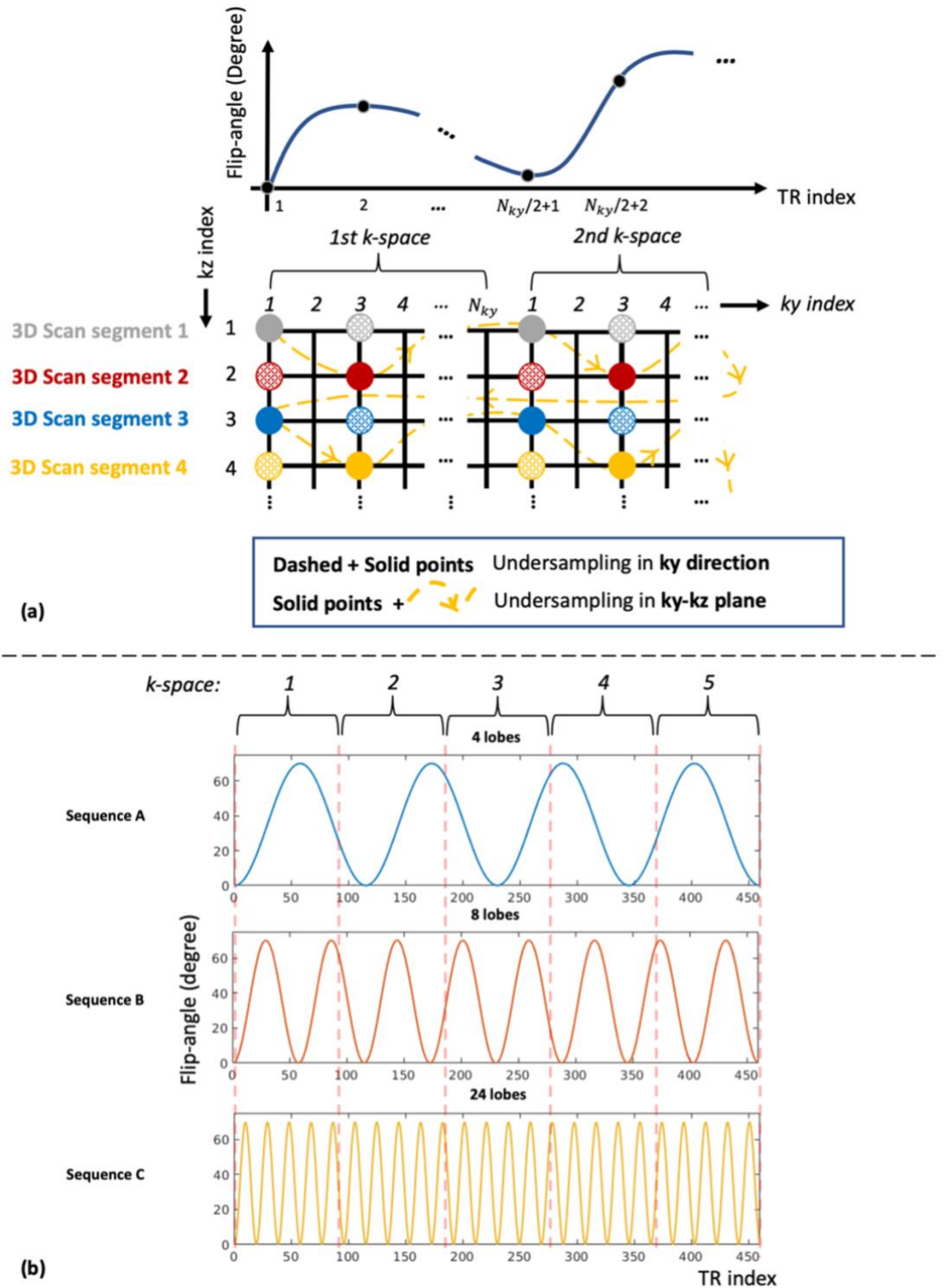

**Figure 2.** Simulated MR signals after multiple repetitions of 3D segments with different $T_w$ and $T_1/T_2$ values. The flip-angle train used is plotted in Figure 1(b) (Sequence A), with $TR = 10.4ms$. Comparing the plots in the two columns, it's shown that different "hyper steady-state" signals will be acquired given different waiting times $T_w$. Few repetitions are required for the simulated signals to reach the "hyper steady-state", and even for the slowest case here in bottom right, only 3 repetitions will be required given the relatively long $T_1, T_2$ and a relatively short $T_w$.

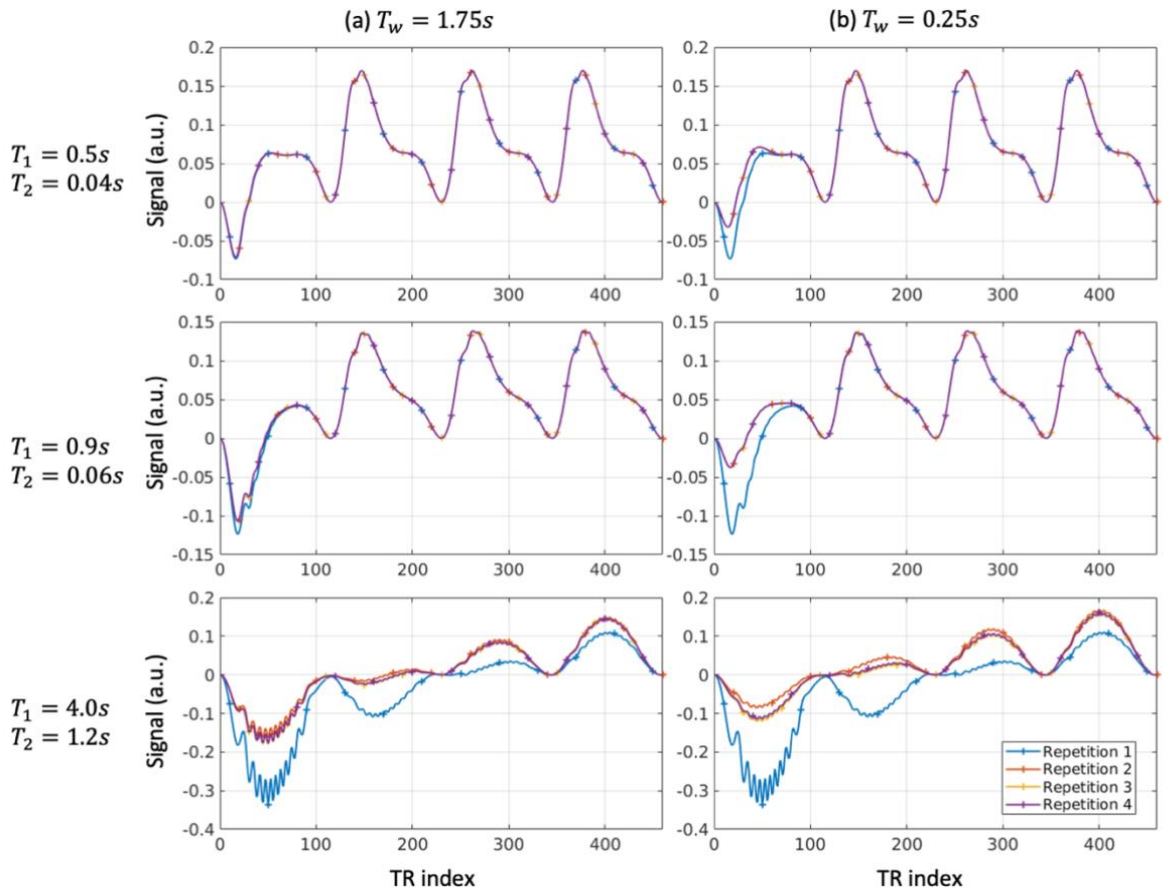

**Figure 3**. Flowchart of the 3D MR-STAT reconstruction with the proposed data decoupling strategy. The whole reconstruction is split into two substeps, as shown in the green and red background boxes. Data in k-space domain are plotted in the left side of the flowchart and image-domain data in the right side. In Substep 1, multi-channel 3D data undersampled in the $ky - kz$ plane is used as the input, and a SENSE reconstruction is run for each of the k-space dataset to achieve fully-sampled 3D k-space data. The unfolded 3D rawdata in image domain is then transformed into k-space along x and y direction, to get 2D 'fully-sampled' MR-STAT data. In Substep 2, $N_z$ separate 2D MR-STAT reconstructions are run.

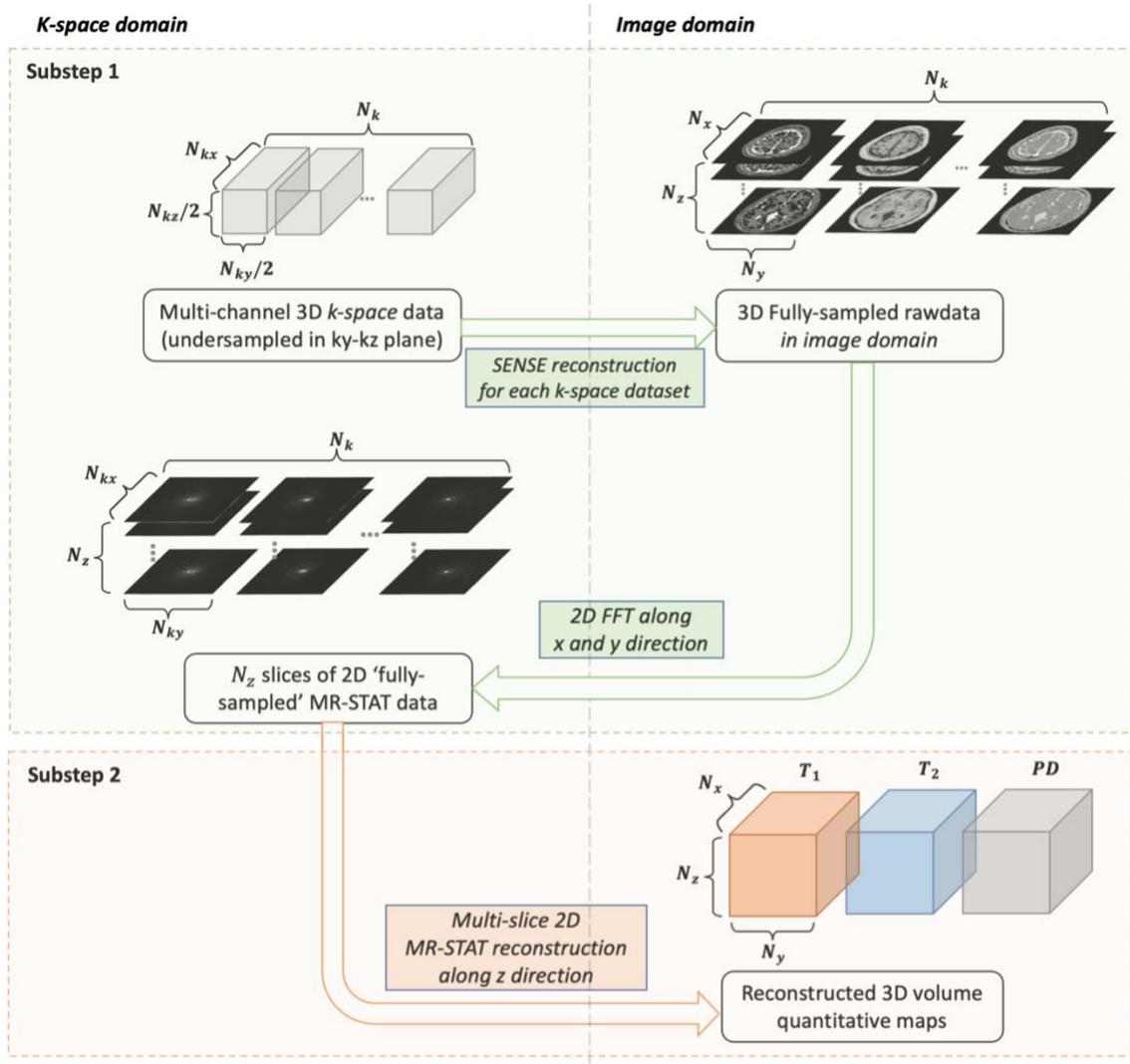

**Figure 4**. Reconstruction results for the digital brain experiment using three different RF flip-angle trains, 4, 8 and 24 sine-square lobes within 5 k-spaces, respectively (See also Figure 1b). Results of the central slice in the 3D volume are shown, and data with undersampling of 4 are used for all 3 experiments. (a) Results of Substep1: SENSE-reconstructed 2D MR-STAT data for each of the 5 k-space after 2D FFT. (b) Results of Substep2: Reconstructed quantitative maps, $T_1$, $T_2$ and $\rho$. The averaged Mean absolute percentage error (MAPE) values for 2D k-space data in (a) are given, and the MAPE values for each quantitative map in (b) are given.

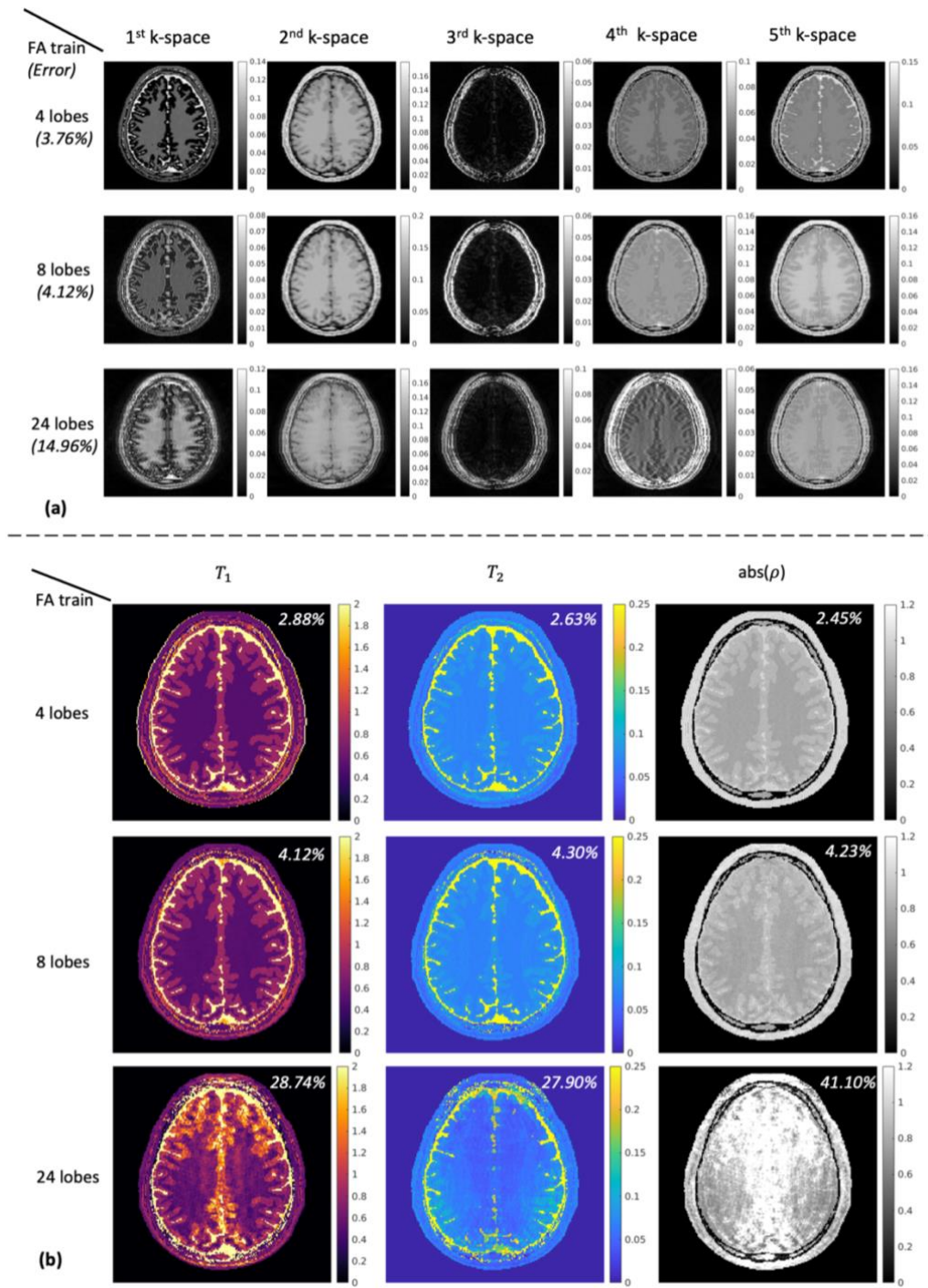

**Table 1.** 3D MR-STAT acquisition parameters for the phantom and volunteer (one-side knee, bilateral lower legs) studies. For phantom and knee experiments, only ky undersampling with an acceleration factor R = 2 is realized on the scanner, and the data are later retrospectively undersampled by another undersampling factor of 2. For the lower leg experiment, ky-kz undersampling with an undersampling factor of 4 is prospectively realized on the scanner. Different coils and different orientations are used for the three experiments. AP: anterior-posterior; LR: left-right; FH: feet-head.

| Scans | Eurospin phantom | Knee | Lower leg |
|---|---|---|---|
| Orientation | Transverse | Sagittal | Coronal |
| Resolution (mm^3) | 1.2 x 1.2 x 1.2 | 0.8 x 0.8 x 1.5 | 1.6 x 1.6 x 1.6 |
| Matrix size | 184 (AP) x 184 (LR) x 100 (FH) | 184 (FH) x 184 (AP) x 100 (LR) | 140 (FH) x 184 (LR) x 100 (AP) |
| FOV (mm^3) | 221 x 221 x 120 | 147 x 147 x 150 | 224 x 294 x 160 |
| Coil | 32-channel head coil | 16-channel T/R knee coil | 28-channel anterior / posterior coil |
| Number of TRs per segment | 460 | 460 | 460 |
| Number of segments | 128 | 128 | 64 |
| Undersampling direction | ky | ky | ky x kz |
| Undersampling factor R | 2 | 2 | 2 x 2 |
| Total scan time on scanner(s) | 13 min 56s | 13 min 56s | 6min 58s |

**Figure 5.** Experimental gel phantom results from the 14-minute (prospectively undersampling on scanner in the ky direction, R = 2) and 7-minute (additional retrospectively undersampling in both ky and kz directions, R = 4) 3D MR-STAT scan. (a) Reconstructed T1 and T2 maps from 7-minute (R=4) 3D MR-STAT scan and the $B_1^+$ map from the B1 dream scan. (b) Mean of T1 and T2 values in 11 different tubes from MR-STAT 3D scan (R= 2 and R = 4) and MR-STAT 2D scan vs gold standard results. (c) Comparison of the sequence efficiency from 2D and 3D (R=2 and 4) MR-STAT results. The sequence efficiency is defined by the $T_1$ or $T_2$ signal-to-noise ratio divided by the square root of scan time per slice and the slice thickness.

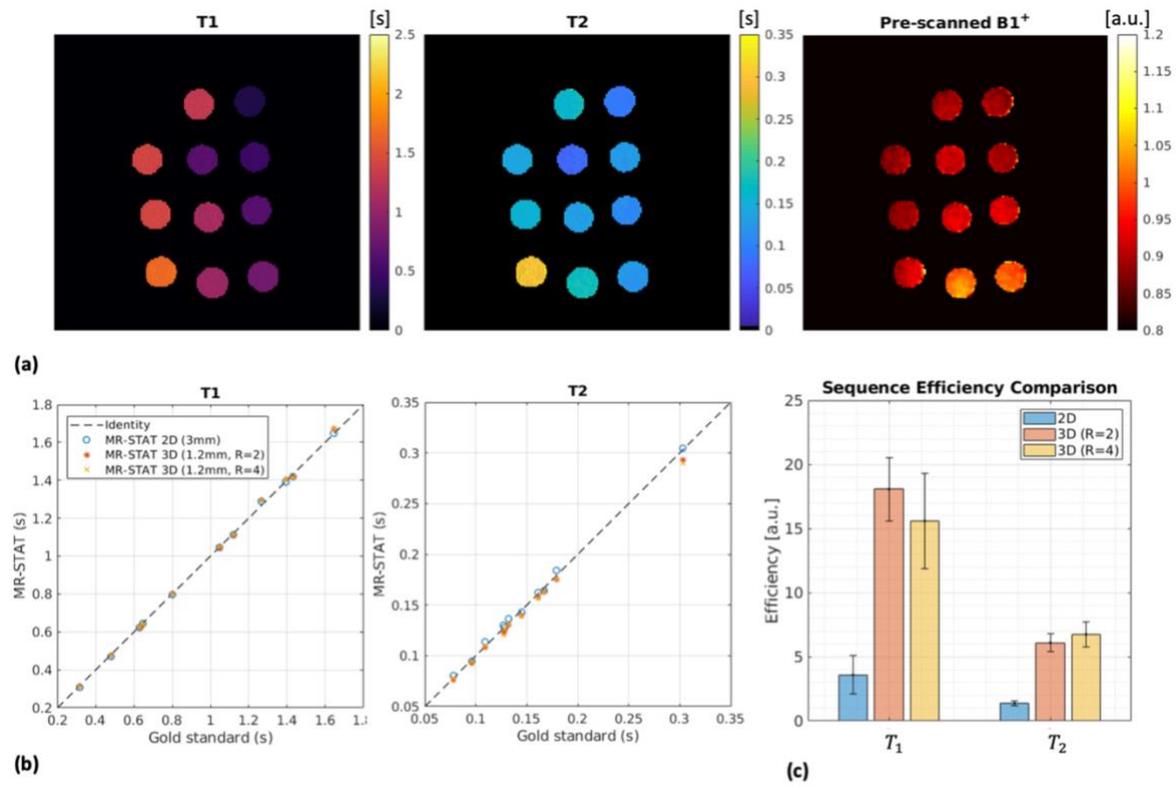

**Figure 6.** Experimental results for the knee scan. Representative (a) sagittal and (b) transverse quantitative maps ($T_1$, $T_2$ and $abs(\rho)$) are shown for both 14-minute ($ky$ undersampling on scanner) and 7-minute (additional retrospective undersampling in $kz$ direction). ROI regions (bone marrow, cartilage, muscle and subcutaneous fat) are drawn on the 3D proton density maps, and mean and std of $T_1$ and $T_2$ values in ROI regions are reported in (c), together with summarized reference values from [25], [34], [35]. Quantitative maps from the 14-min and 7-min sequences show little difference.

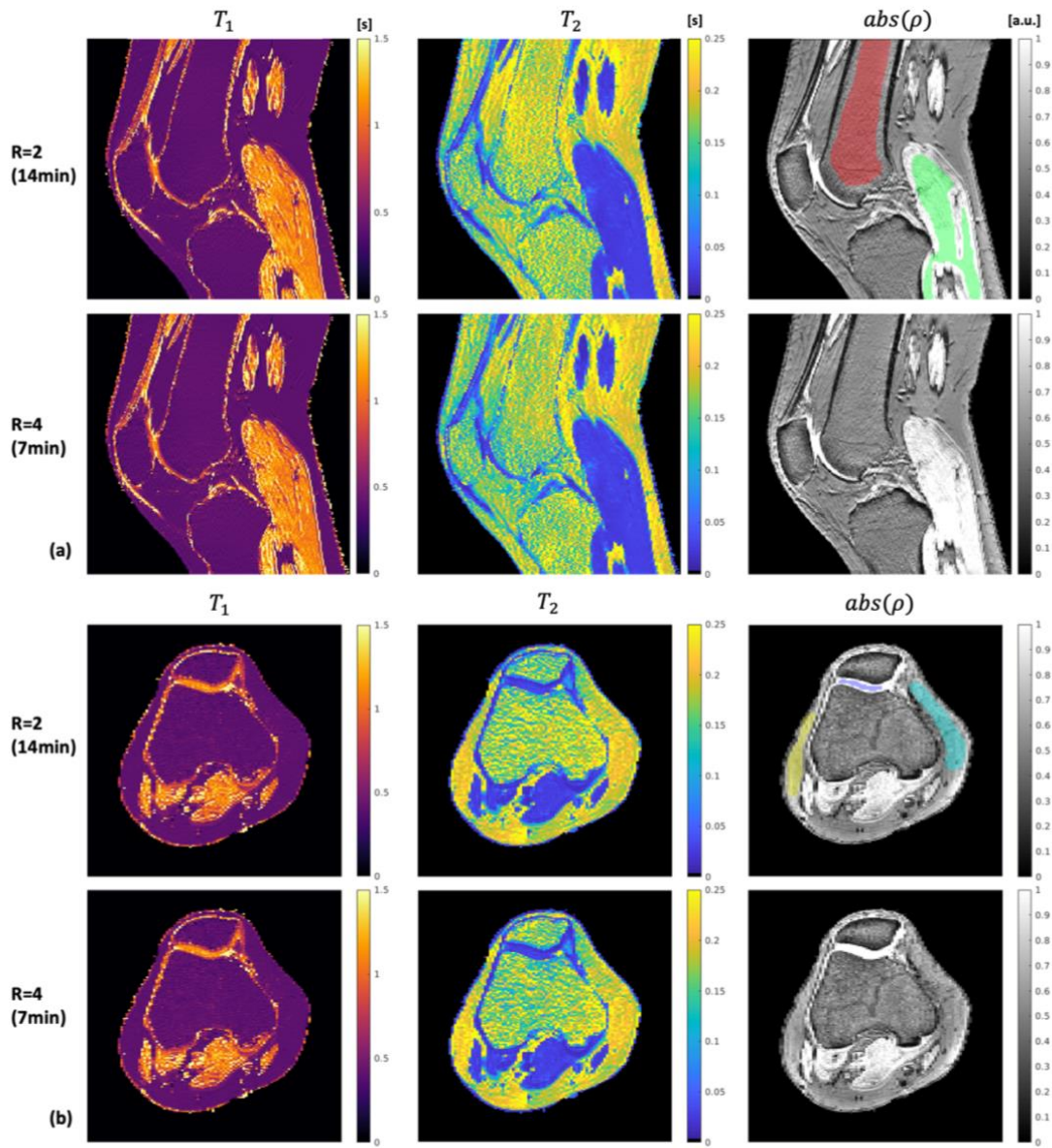

| ROIs | T1 (mean ± std, ms) | | | T2 (mean ± std, ms) | | |
|---|---|---|---|---|---|---|
| | R=2 (14 min) | R=4 (7 min) | Reference range | R=2 (14 min) | R=4 (7 min) | Reference range |
| Region 1 (Bone marrow in (a)) | 409±26.1 | 417±27.2 | 365 – 586 | 202±49.8 | 183±39.7 | 40 – 160 |
| Region 2 (Muscle in (a)) | 1137±175 | 1140±192 | 898 – 1509 | 28.1±8.0 | 29.1±8.9 | 27 – 44 |
| Region 3 (Cartilage in (b)) | 1141±117 | 1144±117 | 778 – 1240 | 29.2±5.7 | 28.2±5.8 | 20 – 45 |
| Region 4 (Fat in (b)) | 411±52.2 | 408±41.4 | 253 – 450 | 202±28.6 | 212±33.9 | 41 – 154 |
| Region 5 (Fat in (b)) | 408±25.8 | 408±25.1 | | 198±28.8 | 196±31.2 | |

(c)

**Figure 7.** Experimental results for the lower leg scan. Representative (a) coronal and (b) transverse quantitative maps ($T_1$, $T_2$ and $abs(\rho)$), as well as the pre-scanned $B_1^+$ maps, are shown for the 7-minute prospective undersampled experiment. Muscle segmentation is manually performed by an experienced researcher and superimposed on the proton density maps. Mean and std of $T_1$ and $T_2$ values in different regions are reported in (c). Arrows in the $B_1^+$ maps show regions with relatively low $B_1^+$ values (about 50% of nominal value), which lead to lower $T_1$ and $T_2$ values in the reconstructed tissue maps.

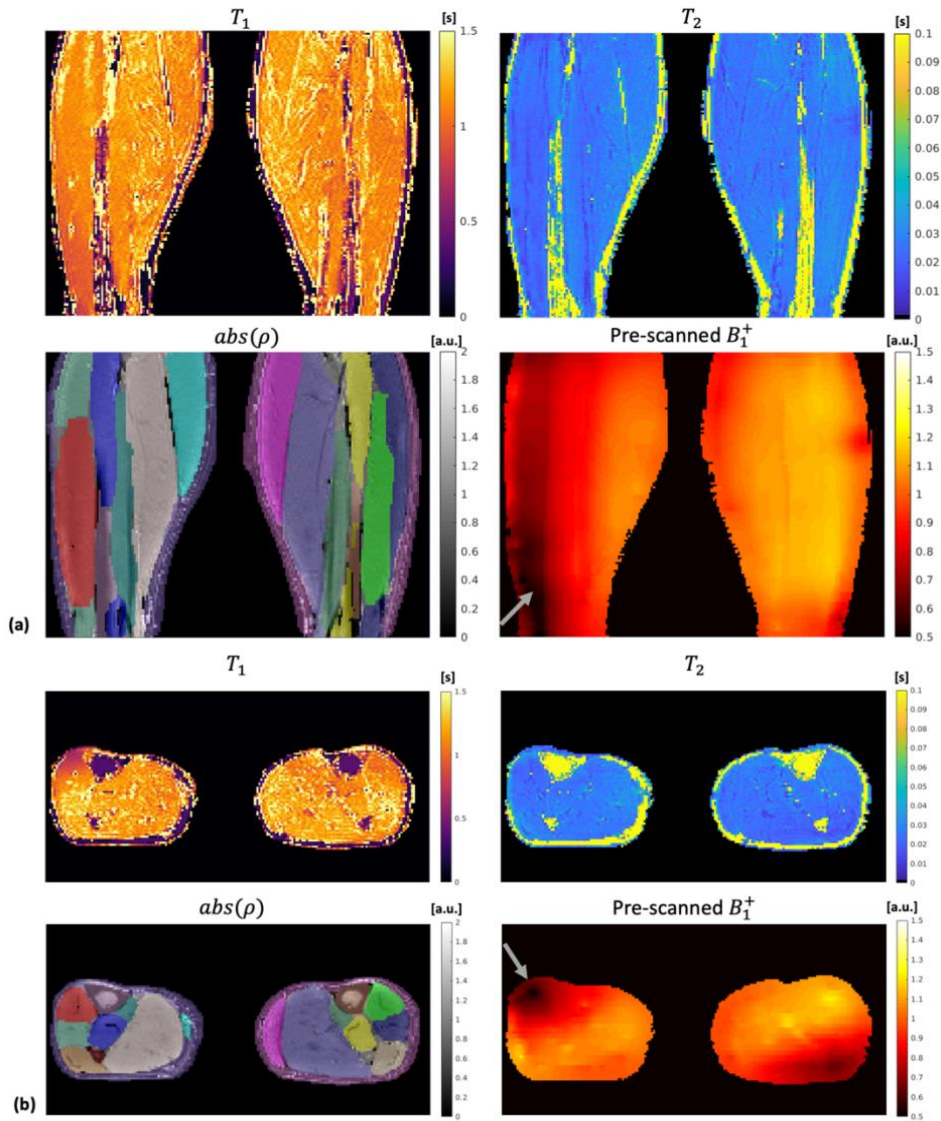

# Supplementary material

**Figure S1.** Reconstructed quantitative maps for the digital brain experiment using fully-sampled 3D MR-STAT data. The fully-sampled data was simulated assuming two continuous $ky$ phase-encoding lines were sampled simultaneously after each RF excitation. Fourier transform along the $kz$ direction was used to acquire 2D MR-STAT data instead of SENSE reconstruction in Substep 1 for each of the multi-channel data. The decoupled 2D data was then reduced to a single virtual channel data via SVD (singular value decomposition) to feed into the 2D MR-STAT reconstruction. These results show that even when using faster oscillating flip-angle trains (8 lobes and 24 lobes), reconstructions can still be run with fully-sampled data; however, the two-step reconstructions fail with the undersampled data as shown in Figure 4.

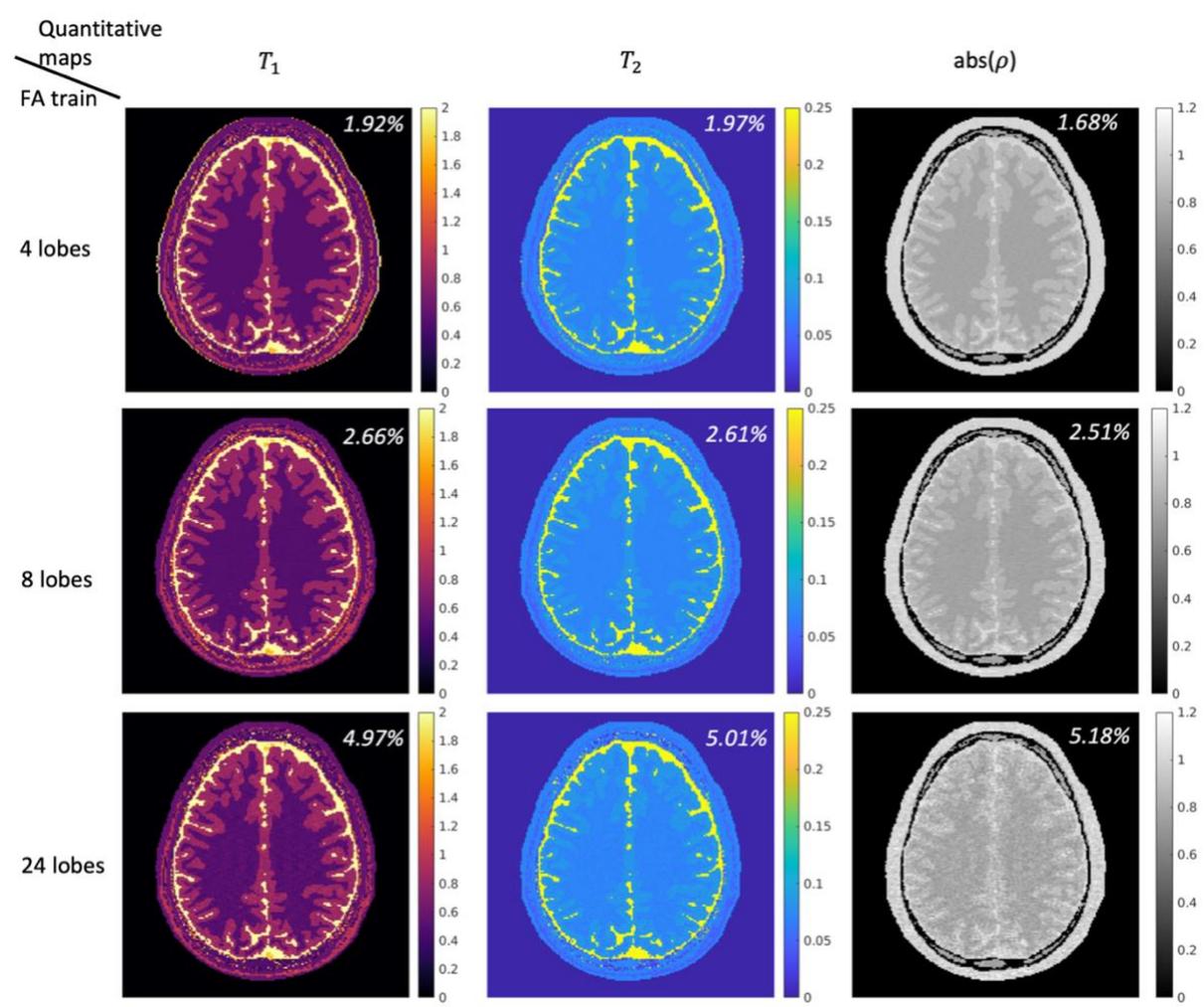

**Figure S2.** Experimental results from brain scan showing CSF flow artifacts. The 3D scan protocol is tested on a brain scan. Transverse (a) and coronal (b) slices (1.2mm$^3$) are shown and compared with 2D MR-STAT results with 3mm slice thickness. Ghosting artifacts originating from the CSF regions can be observed in 3D MR-STAT images, which is most visible along the z direction (feet-head, see black arrows).

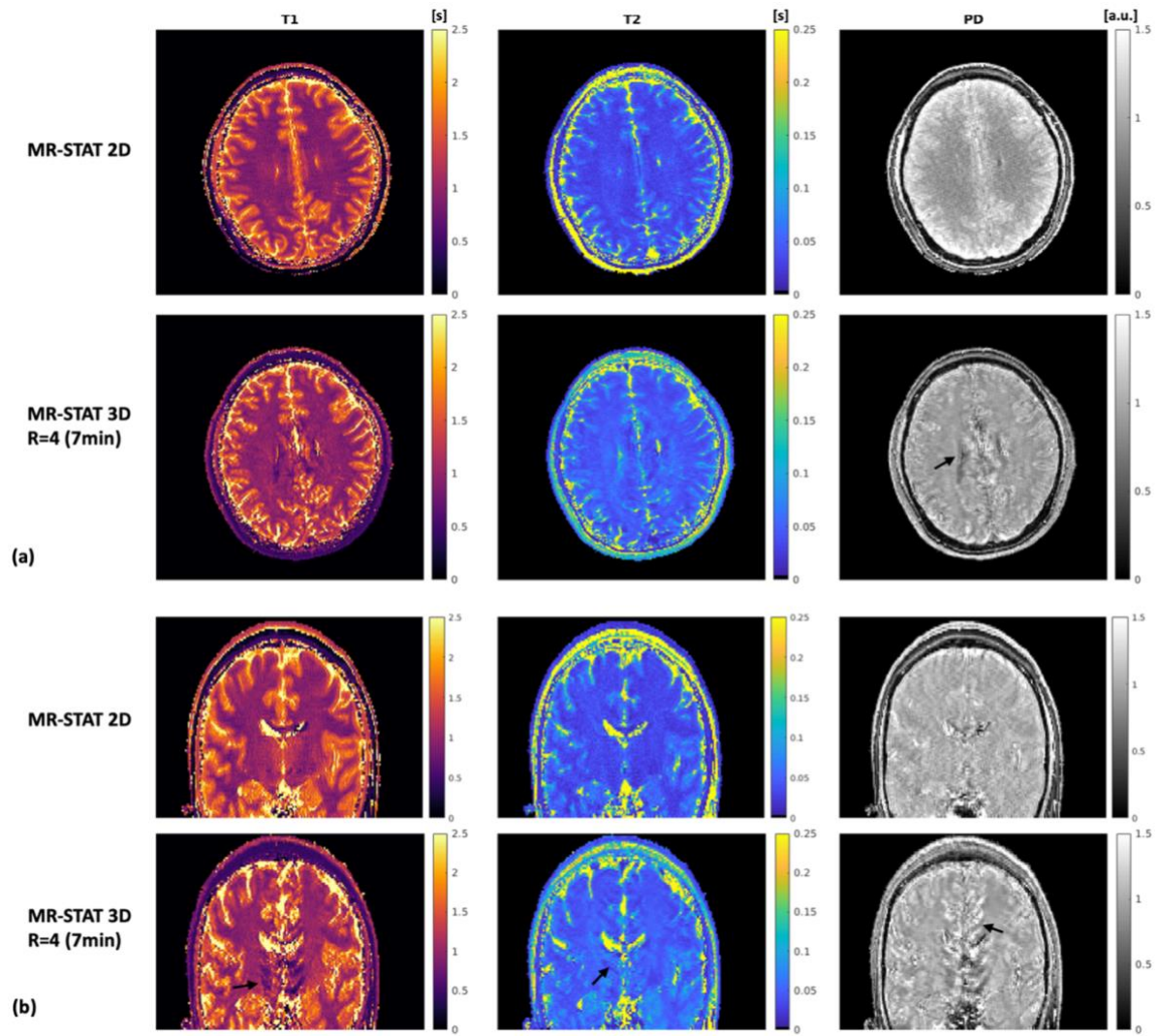